\documentclass[sigplan,screen,nonacm]{acmart}

\usepackage{adjustbox}
\usepackage[ruled,vlined]{algorithm2e}
\usepackage{threeparttable}

\newcommand{\DynaKV}{\textsc{DynaKV}}

\begin{document}

\title{\DynaKV{}: Enabling Accurate and Efficient Long-Sequence LLM Decoding on Smartphones}

\author{Tuowei Wang}
\authornote{Both authors contributed equally to this research.}
\affiliation{%
  \institution{Tsinghua University}
  \city{Beijing}
  \country{China}
}

\author{Minxing Huang}
\authornotemark[1]
\affiliation{%
  \institution{Tsinghua University}
  \city{Beijing}
  \country{China}
}

\author{Fengzu Li}
\affiliation{%
  \institution{Tsinghua University}
  \city{Beijing}
  \country{China}
}

\author{Ligeng Chen}
\affiliation{%
  \institution{Honor Device Co., Ltd.}
  \city{Shenzhen}
  \country{China}
}

\author{Jinrui Zhang}
\affiliation{%
  \institution{Tsinghua University}
  \city{Beijing}
  \country{China}
}

\author{Ju Ren}
\affiliation{%
  \institution{Tsinghua University}
  \city{Beijing}
  \country{China}
}

\begin{abstract}
As the demand for human-like reasoning, multi-turn dialogues, and long-form responses grows, large language models (LLMs) are increasingly expected to support efficient and effective long-sequence decoding. However, due to limited DRAM capacity, long-sequence LLM decoding on smartphones is constrained by the key–value cache (KVCache), whose memory footprint increases linearly with sequence length. Retrieval-based methods mitigate DRAM pressure by offloading KVCache to flash and retrieving query-relevant entries through cluster-based indexing. Unfortunately, as decoding progresses, KVCache distribution shifts render static or local cluster updates progressively misaligned, excluding essential entries or fetching redundant ones. These issues are further exacerbated by smartphone-specific limitations in bandwidth, IOPS, and memory capacity.

We propose \DynaKV{}, the first adaptive KVCache management approach that jointly addresses accuracy and efficiency for long-sequence decoding on smartphones. \DynaKV{} integrates three key techniques: (1) Migration-Free Cluster Adaptation, which adaptively splits clusters during retrieval without incurring additional transfers; (2) Continuity-Centric Flash Management, which co-locates correlated entries and clusters and employs a dual-head layout for efficient updates; and (3) Memory-Efficient Cache Design, which virtualizes cache space across DRAM and flash and extends replacement policies to align with cluster-level access patterns. Evaluations demonstrate that \DynaKV{} improves retrieval accuracy and reduces end-to-end latency compared to state-of-the-art solutions, achieving average gains of $1.38\times$ in accuracy and $1.47\times$ speedups. Furthermore, the insights of \DynaKV{} naturally extend to other long-context workloads and multi-tier memory hierarchies, underscoring its broader applicability.
\end{abstract}

\maketitle

\section{Introduction}
The emergence of large language models (LLMs) has fundamentally transformed the processing of long-context data. Beyond understanding extensive inputs such as documents~\cite{document-1,document-2} or codebases~\cite{codebase-1,codebase-2}, LLMs are increasingly expected to generate coherent and long-form outputs, particularly \textbf{on mobile devices like smartphones}. On one hand, state-of-the-art LLMs are moving toward reasoning-oriented designs~\cite{openai-o1,deepseek-r1,gemini2.5}, leveraging structured thought processes to strengthen problem-solving capabilities. On the other hand, mobile applications such as advanced conversational agents demand both detailed response generation~\cite{longlamp,hellobench,seed-story} and sustaining multi-turn dialogues~\cite{mt-bench,mt-bench++,mt-bench-101}. These trends together underscore the critical need to enable efficient and effective long-sequence LLM decoding on smartphones.

Similar to other long-context scenarios, the limited memory capacity on smartphones poses a major challenge for long-sequence LLM decoding. Since the memory usage of the \textbf{key–value cache (KVCache)} scales linearly with sequence length, it quickly becomes the dominant bottleneck at longer contexts. To mitigate this issue, prior studies have investigated the contextual sparsity inherent in the attention mechanism~\cite{attention-sparsity,deja-vu,jenga,long-exposure}. In practice, each query vector interacts significantly with only a small, query-dependent subset of key and value vectors. Building on this property, a variety of retrieval-based methods~\cite{pqcache,clusterkv,infllm,retrieval-attention} have been proposed for memory-constrained settings, offloading the entire KVCache to external storage and retrieving only the relevant entries into main memory on demand.

\begin{figure}[t]
    \centering
    \includegraphics[width=1.0\linewidth]{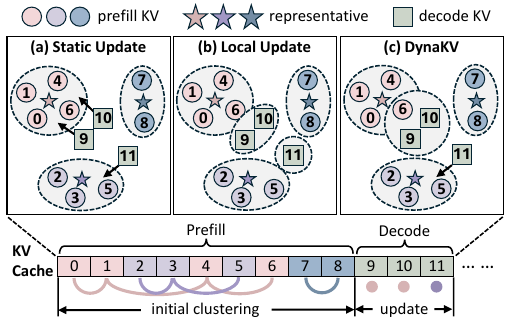}
    \caption{Illustration of three KVCache cluster management strategies. (a) Static Update: new entries are greedily appended to their nearest existing clusters. (b) Local Update: new entries are re-clustered independently of existing clusters. (c) \DynaKV{}: adaptively appends new entries or splits existing clusters to maintain accuracy and efficiency.}
    \label{fig:comparison}
\end{figure}

To balance retrieval efficiency and accuracy, the KVCache is typically partitioned into clusters based on embedding similarity, which then serve as the fundamental retrieval units. Since global clustering is both computationally and memory-intensive, it is performed only once during the prefill phase. Newly generated KV entries are either greedily appended to the nearest existing clusters (\textit{Static Update})~\cite{pqcache,squeezed-attention} or periodically re-clustered within a local context window (\textit{Local Update})~\cite{clusterkv,shadowkv}, as illustrated in Figure~\ref{fig:comparison}. These methods remain effective when the prefill proportion is relatively large. However, the situation changes in long-sequence decoding, where generated tokens dominate the inference context.

Specifically, we observe that as the decode length increases, the distribution of the KVCache undergoes continual transformation, exhibiting clustering patterns that diverge substantially from the initial ones, a phenomenon we term \textbf{KVCache Distribution Shift}. Unfortunately, existing clustering methods fail to effectively accommodate this dynamic shift, resulting in suboptimal KVCache retrieval. In particular, excluding essential entries degrades inference accuracy, while transferring redundant entries wastes I/O bandwidth. This observation motivates a critical question: \textit{Can we develop an adaptive KVCache management that preserves both accuracy and efficiency of KVCache retrieval on smartphones?}

However, this is not a low-hanging fruit, as it stems from the inherent misalignment between the adaptive demands of clustering mechanisms and the stringent constraints of smartphone memory hierarchies. The presence of KVCache clustering shift necessitates access to global information for adaptive clustering, which unavoidably incurs additional data transfers from external storage to main memory. Given the hardware characteristics of smartphones, such I/O operations are readily bottlenecked by both limited \textbf{bandwidth}, restricted \textbf{Input/Output Operations Per Second (IOPS)}, and constrained \textbf{memory capacity}. Our analysis highlights three critical technical challenges that must be tackled:

\noindent(1) \textbf{Bandwidth-Limited Adaptation.} Adaptive clustering requires continuous updates to existing clusters as new KV entries are generated. Since external storage lacks compute capability, the relevant clusters must first be transferred into main memory before being updated. The resulting frequent transfers can quickly lead to I/O performance bottlenecks.

\noindent(2) \textbf{IOPS-Bounded Flash Access.} Retrieved KV entries exhibit scattered and dynamic access patterns across different queries. Placing the KVCache naively in strict sequence order results in fragmented flash accesses, exacerbating the IOPS bottleneck. Alternatively, organizing it by clusters improves continuous reads but raises challenges for frequent updates.

\noindent(3) \textbf{Memory-Constrained Caching.} Retaining frequently retrieved KV entries in main memory is essential for reducing transfer overhead. However, the distinctive access patterns of KVCache render standard replacement policies ineffective. Moreover, cluster-based organization inflates cache demands, which is incompatible with the limited memory capacity.

In this paper, we propose \textbf{\DynaKV{}}, an efficient approach that enables long-sequence LLM decoding on smartphones with adaptive KVCache management. \DynaKV{} employs three key techniques to tackle these challenges, respectively:

\noindent(1) \textbf{Migration-Free Cluster Adaptation.} \DynaKV{} design a variance-based self-adaptive clustering algorithm that preserves clustering effectiveness while maintaining cluster granularity as coarse as possible. To avoid extra I/O overhead, \DynaKV{} adopts a delayed-split strategy that integrates cluster splitting naturally into the existing retrieval workflow.

\noindent(2) \textbf{Continuity-Centric Flash Management.} To facilitate continuous read access and mitigate IOPS bottlenecks, \DynaKV{} places all entries within a cluster as well as highly-correlated clusters continuously in flash. To support frequent adaptations, \DynaKV{} employs a dual-head layout that allows cluster updates without interfering with each other.

\noindent(3) \textbf{Memory-Efficient Cache Design.} \DynaKV{} constructs a virtual cache space that unifies DRAM and flash. By leveraging idle bandwidth during computation, cache clusters are seamlessly swapped between the two tiers. To further improve cache hit rate, \DynaKV{} augments standard replacement policies to align with cluster-based retrieval.

We evaluate \DynaKV{} on three smartphones with distinct hardware configurations, benchmarking a diverse range of LLMs varying in structures and scales. The results demonstrate that \DynaKV{} outperforms the state-of-the-art solutions in retrieval accuracy during long-sequence decoding, achieving improvements of $1.38\times$ on average. Moreover, the system-level optimizations guarantee retrieval efficiency, delivering average speedups of $1.47\times$ in end-to-end latency.

To the best of our knowledge, \DynaKV{} is the first to simultaneously achieve both accuracy and efficiency of long-sequence LLM decoding on smartphones. Our main contributions can be summarized as follows:
\begin{itemize}
    \item We identify the phenomenon of KVCache distribution shift in long-sequence LLM decoding, which degrades both the accuracy and efficiency of KVCache retrieval.
    \item We develop three core techniques for adaptive KVCache management that integrate algorithmic innovations with system-level optimizations, overcoming the constraints of smartphone memory hierarchies.
    \item We conduct extensive evaluations across various LLMs and mobile hardware, demonstrating substantial improvements over state-of-the-art solutions.
\end{itemize}

\begin{table}[b]
    \small
    \centering
    \caption{KVCache memory footprint (relative to 8-bit model weights) of Llama3.2~\cite{llama3.2} across sequence lengths $s$.}
    \label{tab:kvcache-memory}
    \begin{adjustbox}{width=1.0\linewidth,center}
    \setlength{\tabcolsep}{3pt}
    \begin{tabular}{l|c|cccc}
        \toprule
        \textbf{Model}  & \textbf{Weights} & $s=8K$ & $s=16K$ & $s=32K$ & $s=64K$ \\ \midrule
        Llama3.2-1B     & 1.2 GB           & 0.3 GB & 0.5 GB  & 1.1 GB  & 2.1 GB  \\
        KVCache/Weights & -                & 25.0\% & 41.7\%  & 91.7\%  & 175.0\% \\ \midrule
        Llama3.2-3B     & 3.2 GB           & 0.9 GB & 1.9 GB  & 3.8 GB  & 7.5 GB  \\
        KVCache/Weights & -                & 28.1\% & 59.4\%  & 118.8\% & 234.4\% \\
        \bottomrule
    \end{tabular}
    \end{adjustbox}
\end{table}

\section{Background and Motivation}
\subsection{KVCache Retrieval in LLM Inference}\label{sec:kvcache-retreival}
Almost all state-of-the-art LLMs adopt a transformer-based architecture that generates outputs in an autoregressive manner. During LLM inference, tokens are generated sequentially, with each new token attending not only to the most recent hidden states but also to all preceding tokens in the sequence. To avoid recomputing the hidden states of earlier tokens, the model typically stores the intermediate key and value vectors from each attention layer in a dedicated KVCache.

While the KVCache greatly reduces computation during inference, it incurs a substantial memory footprint, particularly in long-context scenarios where the cache must grow linearly with sequence length. As shown in Table~\ref{tab:kvcache-memory}, the KVCache exceeds even the model weights in size at longer context, contributing 64\%-70\% of the total memory footprint. To address this, recent studies~\cite{attention-sparsity,deja-vu,jenga} explore the inherent \textbf{Contextual Sparsity} of attention: for computing the next token, only a subset of the stored keys and values significantly contribute to the attention output, and the relevant subset varies dynamically with individual tokens.

\begin{figure}[t]
    \centering
    \includegraphics[width=1.0\linewidth]{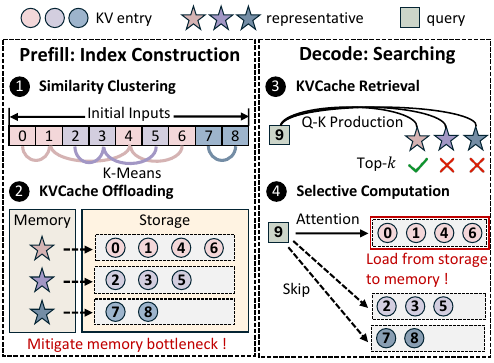}
    \caption{Two-phase KVCache retrieval process. Offloading KVCache from memory to storage mitigates memory bottlenecks, but shifts the critical bottleneck to I/O operations.}
    \label{fig:kv-retrieval}
\end{figure}

This sparsity property enables \textbf{Targeted Retrieval} on the KVCache~\cite{pqcache,clusterkv,infllm,retrieval-attention}, where only the most relevant KV entries for a given query are accessed for processing, analogous to information retrieval in database systems~\cite{anns-1,anns-2,anns-3}. Specifically, the KVCache can be offloaded to external storage rather than maintained entirely in main memory, with relevant entries transferred back on demand. Figure~\ref{fig:kv-retrieval} illustrates a typical two-phase KVCache retrieval process:

\noindent(1) \textbf{Prefill Phase:} The KVCache is partitioned into clusters based on embedding similarity and stored in external storage. Each cluster is assigned a representative to facilitate indexing. These clusters then serve as the fundamental retrieval units, balancing accuracy and efficiency in subsequent retrieval.

\noindent(2) \textbf{Decode Phase:} Given a newly generated query, the top-$k$ relevant KV clusters are identified by comparing the query against the cluster representatives. This paradigm effectively mitigates the KVCache memory bottleneck, particularly on resource-constrained devices such as smartphones.

\subsection{Two-Tier Memory Hierarchy on Smartphones}
Mobile Devices such as smartphones typically employ a two-tier memory hierarchy, as illustrated in Figure~\ref{fig:memory-hierarchy}(a). The first tier is \textbf{Dynamic Random-Access Memory DRAM)}, which provides fast access and supports direct computation. Due to the cost and power constraints, the size of DRAM is relatively small, typically only a few gigabytes (GB). The second tier is \textbf{Flash Memory}, primarily used for external storage and commonly accessed through universal flash storage (UFS)~\cite{ufs}. Leveraging NAND flash provides much larger capacity than DRAM at a lower cost, with scalability reaching terabyte (TB) levels. However, since flash memory lacks compute capability, data must first be transferred into DRAM before computation can be performed. Our comprehensive evaluation shows that this asymmetric two-tier hierarchy introduces three key challenges for KVCache retrieval on smartphones:

\begin{figure}[t]
    \centering
    \includegraphics[width=0.95\linewidth]{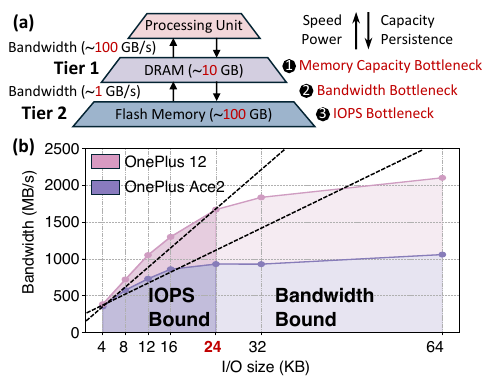}
    \caption{(a) Two-tier memory hierarchy on smartphones (DRAM and flash), highlighting three key bottlenecks. (b) UFS bandwidth under various continuous I/O sizes, where the limiting factor shifts from IOPS to bandwidth capacity.}
    \label{fig:memory-hierarchy}
\end{figure}

\noindent(1) \textbf{Bandwidth.} Retrieval-based paradigm requires transferring relevant KVCache from external storage into memory at each inference step. Consequently, the I/O bandwidth between flash and DRAM becomes a fundamental constraint. While DRAM can sustain hundreds of GB/s of throughput, the bandwidth of flash memory remains relatively low (e.g., 2.9 GB/s per lane in the latest UFS 4.0). Consequently, I/O transfers often dominate inference latency, with data transfer overhead emerging as a primary performance bottleneck.

\noindent(2) \textbf{IOPS.} Unlike server-grade storage such as NVMe, UFS on smartphones typically supports only a shallow command queue (e.g., 32 entries). This constraint reduces the achievable IOPS for flash reads and can even prevent full utilization of the available I/O bandwidth. As illustrated in Figure~\ref{fig:memory-hierarchy}(b), the read bandwidth increases with the size of continuous I/O, since larger continuous reads can be issued by a single read operation. Specifically, when the continuous I/O size is below 24 KB, the bandwidth scales almost linearly with the I/O size, indicating that these reads are primarily IOPS-bound.

\noindent(3) \textbf{Memory Capacity.} The limited capacity of DRAM further exacerbates the above challenges. Since the KVCache grows linearly with sequence length, it can quickly exceed available memory even for moderate contexts. Cluster-based KVCache organization amplifies memory demand, as entire clusters must be managed as retrieval units. On one hand, this constraint forces frequent offloading to flash and intensifies dependence on bandwidth- and IOPS-constrained I/O. On the other hand, it underscores the need for efficient cache design to maximize the utility of scarce DRAM resources.

\begin{figure}[t]
    \centering
    \includegraphics[width=1.0\linewidth]{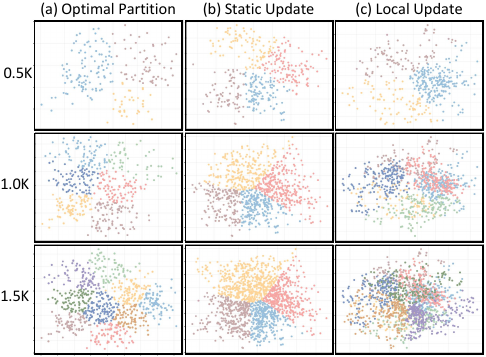}
    \caption{Visualization of KVCache distribution shift during decoding using PCA~\cite{pca}, with Qwen2.5-3B evaluated on the MMLU dataset~\cite{mmlu}. Different colors denote distinct clusters.}
    \label{fig:distribution-shift}
\end{figure}

\subsection{Motivation: KVCache Distribution Shift}
The cluster-based organization of the KVCache improves retrieval efficiency by enabling queries to be compared only against cluster representatives rather than individual entries. However, the effectiveness of this approach critically depends on clustering quality. As the similarity among entries within a cluster decreases, its representative becomes less informative. This misalignment can lead to degraded inference accuracy when essential entries are excluded, and wasted I/O bandwidth when redundant entries are retrieved.

Theoretically, the optimal partition of the KVCache can be achieved through global clustering over the entire cache. However, because global clustering is both computationally and memory-intensive, it is typically performed only once during the prefill phase for initial partition. During decoding, newly generated KV entries are then managed using one of two strategies: (1) \textbf{Static Update}, which greedily assigns each new entry to its nearest existing cluster~\cite{pqcache,squeezed-attention}, or (2) \textbf{Local Update}, which re-clusters new entries separately from the existing KVCache~\cite{clusterkv,shadowkv}. Guided by the principle of minimally modifying the initial partition, both strategies introduce little update overhead and remain effective when the prefill proportion dominates the inference context.

In this paper, we observe an unexpected shift in KVCache distribution during decoding. As shown in Figure~\ref{fig:distribution-shift}(a), the optimal partition (i.e., global clustering over the entire KVCache) progressively diverges from the initial partition as the decode length grows. Furthermore, existing strategies fail to adapt effectively to this shift. (1) Figure~\ref{fig:distribution-shift}(b) shows that static update enlarges cluster sizes and raises intra-cluster variance. Under the KVCache distribution shift, newly generated entries may be distant from all existing clusters, making greedy appending increasingly ineffective. (2) Figure~\ref{fig:distribution-shift}(c) shows that local update fragments the KVCache by generating many small clusters. Without global information, these clusters may fail to align with the overall KVCache distribution. This fragmentation also results in fine-grained read accesses, which are more likely to hit IOPS limitations.

This phenomenon is particularly pronounced in \textbf{Long-Sequence Decoding} scenarios, where the decode length far exceeds the prefill length. With the growing demand for human-like reasoning~\cite{openai-o1,gemini2.5,deepseek-r1}, multi-turn conversations~\cite{mt-bench,mt-bench++,mt-bench-101}, and long-form response generation~\cite{longlamp,hellobench,seed-story}, the need for more adaptive and efficient KVCache management has become critical, yet remains largely underexplored.

\begin{figure}
    \centering
    \includegraphics[width=0.95\linewidth]{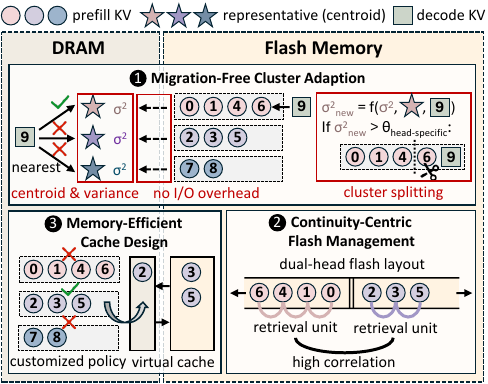}
    \caption{Overview of \DynaKV{}.}
    \label{fig:overview}
\end{figure}

\section{System Overview}
We propose \DynaKV{}, an efficient approach that enables long-sequence LLM decoding on smartphones with adaptive KVCache management. Unlike prior work that primarily targets KVCache memory bottlenecks, \DynaKV{} highlights the dynamic shifts in KVCache distribution during decoding. By combining algorithmic innovations with system-level design, \DynaKV{} achieves an effective balance between accuracy and efficiency. Figure~\ref{fig:overview} shows an overview of \DynaKV{}.

\noindent\textbf{Migration-Free Cluster Adaptation (\S~\ref{sec:cluster}).} \DynaKV{} begins with an algorithm that adaptively updates KVCache clusters with minimal data migration. Based on initial partition, \DynaKV{} tracks the intra-cluster variance as an effectiveness metric and triggers a split once it exceeds a head-specific threshold. Particularly, splits are applied only to in-memory clusters and deferred for disk-resident clusters, eliminating extra I/O overhead while preserving correctness.

\noindent\textbf{Continuity-Centric Flash Management (\S~\ref{sec:flash}).} Given the IOPS bottlenecks, \DynaKV{} optimizes the KVCache placement on flash to facilitate continuous read access. Specifically, \DynaKV{} leverages the co-retrieval property of KVCache by storing all entries within a cluster as well as highly-correlated clusters contiguously. \DynaKV{} further adopts a dual-head layout that places a pair of clusters back-to-back, enabling frequent cluster updates without costly data permutation.

\noindent\textbf{Memory-Efficient Cache Design (\S~\ref{sec:cache}).} \DynaKV{} designs a virtual cache space to mitigate the limited memory capacity. By leveraging idle bandwidth during computation, a portion of cached entries is offloaded to flash and seamlessly fetched back on demand. \DynaKV{} also customizes cache replacement policies to align with cluster-based retrieval. By exploiting the structure and dynamics of retrieval patterns, small and recently updated ones are assigned higher priority.

\section{Migration-Free Cluster Adaptation}\label{sec:cluster}
\subsection{Problem Formulation}
\noindent\textbf{Long-Sequence Decoding.} We consider a typical two-phase process for LLM inference. In the prefill phase, the model processes an input prompt of length $L_{0}$ and generates the initial KVCache $\mathcal{K}_{0}$:
\begin{equation}
    \mathcal{K}_{0}=\{\mathbf{k}_{\text{init}}^{(1)},\mathbf{k}_{\text{init}}^{(2)},...,\mathbf{k}_{\text{init}}^{(L_{0})}\}
\end{equation}
where each $\mathbf{k}_{i}$ denotes a key-value (KV) vector pair.

Following the prefill phase, the model enters the decoding phase, where tokens are generated autoregressively. At each step $t$, the model produces a new KV pair $\mathbf{k}_{\text{new}}^{(t)}$. The accumulated KVCache up to step $t$ can thus be expressed as:
\begin{equation}
    \mathcal{K}_{t}=\mathcal{K}_{0}\cup\{\mathbf{k}_{\text{new}}^{(1)},\mathbf{k}_{\text{new}}^{(2)},...,\mathbf{k}_{\text{new}}^{(t)}\}
\end{equation}

In this paper, we focus on long-sequence decoding, which is prevalent in scenarios such as enhanced reasoning and multi-turn dialogues. In these cases, the total number of decoded tokens substantially exceeds the initial prefill length.

\noindent\textbf{Cluster-Based Retrieval.} As discussed in Section~\ref{sec:kvcache-retreival}, KVCache $K_{t}$ is partitioned into $M_{t}$ disjoint clusters using algorithms such as K-Means and stored on flash:
\begin{equation}
    \mathcal{P}_{t}=\{\mathcal{C}_{1}^{(t)},\mathcal{C}_{2}^{(t)},...,\mathcal{C}_{M_{t}}^{(t)}\},\quad \bigcup_{j=1}^{M_{t}}\mathcal{C}_{j}^{(t)}=\mathcal{K}_{t}
\end{equation}
where each $\mathcal{C}{j}^{(t)}$ denotes a KVCache cluster.

For each cluster $\mathcal{C}{j}^{(t)}$, a representative $r_{j}^{(t)}$ is selected, typically chosen as the centroid of the cluster:
\begin{equation}
    r_{j}^{(t)} = \frac{1}{\left|\mathcal{C}{j}^{(t)}\right|} \sum_{\mathbf{k}\in \mathcal{C}_{j}^{(t)}} \mathbf{k},
\end{equation}
where $\left|\mathcal{C}_{j}^{(t)}\right|$ denotes the number of KV pairs in cluster $\mathcal{C}_{j}^{(t)}$.

These representatives are stored in DRAM and serve as compact proxies for their respective clusters during retrieval. For a given query $q$, the similarity scores between $q$ and all representatives are computed. The top-$k$ clusters are then selected, referred to as the active set $\mathcal{P}_{\text{req}}^{(t)}\subseteq\mathcal{P}_{t}$, transferred back from flash to DRAM for further computation.

\noindent\textbf{Optimization Objectives.} The presence of KVCache distribution shift indicates that the optimal partition $\mathcal{P}_{t}$ at decoding step $t$ diverges increasingly from the initial partition $\mathcal{P}_{0}$. The goal of \DynaKV{} is to design an adaptive clustering algorithm that jointly optimizes the following three objectives:

\noindent(1) \textbf{Accurate Coverage of Required Entries.} Let $\mathcal{K}_{\text{req}}^{(t)}\subseteq\mathcal{K}_{t}$ denote the exact set of KV entries required at decoding step $t$. Excess entries waste I/O bandwidth, while insufficient entries degrade inference accuracy. Therefore, the selected clusters $\mathcal{P}_{\text{req}}^{(t)}$ should approximate $\mathcal{K}_{\text{req}}^{(t)}$ as closely as possible:
\begin{equation}
    \bigcup_{\mathcal{C}\in\mathcal{P}_{\text{req}}^{(t)}}\mathcal{C}\approx\mathcal{K}_{\text{req}}^{(t)}
\end{equation}

\noindent(2) \textbf{Coarse Granularity of Cluster Partition.} Subject to accuracy, the cluster partition should be as coarse as possible. Fewer clusters reduce the number of representatives, improving retrieval efficiency. Also, overly fragmented clusters leave the system IOPS-bound, leading to I/O inefficiency:
\begin{equation}
    \text{minimize}\ \left|\mathcal{P}_{\text{req}}^{(t)}\right|
\end{equation}

\noindent(3) \textbf{Low Overhead from Cluster Updates.} For efficiency, runtime adaptation of cluster partitions should introduce only minimal latency. Since I/O is the primary bottleneck on smartphones, the crucial requirement is to keep the additional I/O load from cluster updates as low as possible:
\begin{equation}
    \text{minimize}\ \text{Load}_{\text{io}}(\mathcal{P}_{t},\mathcal{P}_{\text{req}}^{(t)})
\end{equation}

\begin{figure}[t]
    \centering
    \includegraphics[width=1.0\linewidth]{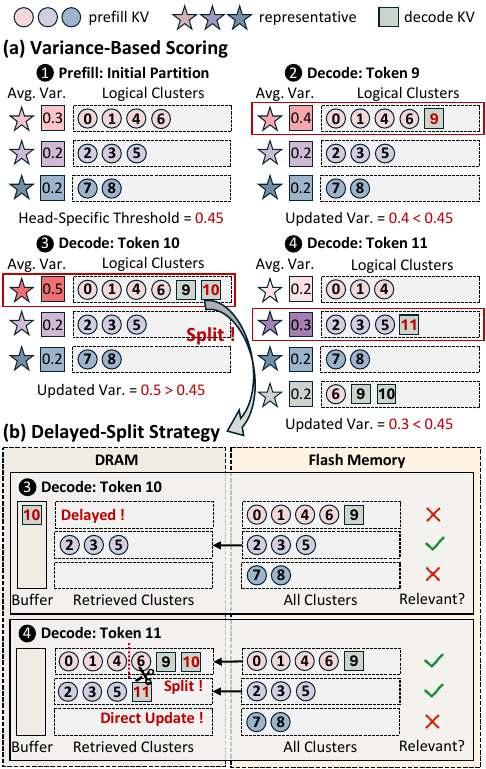}
    \caption{Example of adaptive KVCache management in \DynaKV{}. (a) Intra-cluster variance is monitored to maintain clustering effectiveness. (b) Cluster splitting is integrated into the retrieval workflow, avoiding extra I/O overhead.}
    \label{fig:cluster-adaptation}
\end{figure}

\begin{algorithm}[t]
\small
\DontPrintSemicolon
\caption{Adaptive KVCache Management}
\label{alg:cluster-adaptation}
\KwIn{Prefill KVCache $\mathcal{K}_0$, head-specific thresholds $\{\tau_{h}\}$, buffer budget $B_{\max}$.}

\tcp{Prefill: build the initial partition}
Initial partition $\mathcal{P}_0 \gets \textsc{NewCluster}(\mathcal{K}_0)$\;
\ForEach{cluster $\mathcal{C}_j \in \mathcal{P}_0$}{
  Initialize cluster size $n_j$, centroid $r_j$, variance $\sigma_j^2$\;
  Set buffer $\mathsf{Buf}_j \gets \emptyset$ and flag $f_j \gets \emptyset$\;
}

\tcp{Decode: adaptively update partition}
\For{decoding step $t = 1,2,\ldots$}{
  Load top-$k$ clusters $\mathcal{P}^{(t)}_{\mathrm{req}}$ by query–centroid similarity\;
  Assign new KV entry $\mathbf{k}^{(t)}_{\mathrm{new}}$ to its nearest cluster index $j$\;
  Update $\sigma_j^2 \gets \textsc{UpdateVar}(n_j, r_j, \sigma_j^2, \mathbf{k}^{(t)}_{\mathrm{new}})$\;
  \uIf{$\sigma_j^2 \le \tau_{h}$}{
    Update $\mathcal{C}_j \gets \textsc{AddEntry}(\mathcal{C}_j, \mathbf{k}^{(t)}_{\mathrm{new}}$)\;
    Update $n_j, r_j \gets \textsc{UpdateStats}(n_j, r_j, \mathcal{C}_j)$\;
  }
  \ElseIf{$\mathcal{C}_j \in \mathcal{P}^{(t)}_{\mathrm{req}}$}{
    Update $\mathcal{P}_t \gets \textsc{SplitCluster}(\mathcal{P}_t, \mathcal{C}_j, \mathbf{k}^{(t)}_{\mathrm{new}})$\;
    Update $n_j, r_j \gets \textsc{UpdateStats}(n_j, r_j, \mathcal{C}_j)$\;
  }
  \Else{
    Buffer $\mathbf{k}^{(t)}_{\mathrm{new}}$ in $\mathsf{Buf}_j$ and mark $f_j \gets \textsc{Split}$\;
  }
  
  \tcp{Delayed cluster splitting}
  \ForEach{$\mathcal{C}_j \in \mathcal{P}^{(t)}_{\mathrm{req}}$ with $f_j = \textsc{Split}$}{
    Update $\mathcal{P}_t \gets \textsc{SplitCluster}(\mathcal{P}_t, \mathcal{C}_j, \mathsf{Buf}_j)$\;
    Reset $\mathsf{Buf}_j, f_j \gets \emptyset$\;
  }
  \If{$\sum_j |\mathsf{Buf}_j| \ge B_{\max}$}{
    Load cluster $\mathcal{C}_{j^\dagger}$ with $j^\dagger \gets \arg\max_j |\mathsf{Buf}_j|$\;
    Update $\mathcal{P}_t \gets \textsc{SplitCluster}(\mathcal{P}_t, \mathcal{C}_{j^\dagger}, \mathsf{Buf}_{j^\dagger})$\;
    Reset $\mathsf{Buf}_{j^\dagger}, f_{j^\dagger} \gets \emptyset$\;
  }
}
\end{algorithm}

\subsection{Self-Adaptive Clustering Algorithm}
As illustrated in Figure~\ref{fig:cluster-adaptation}, \DynaKV{} employs a self-adaptive clustering algorithm to jointly satisfy these objectives. The algorithm is composed of two key components: a variance-based scoring mechanism that preserves clustering effectiveness, and a delayed-split strategy that ensures efficient cluster updates with minimal I/O overhead.

\noindent\textbf{Variance-Based Scoring.} As outlined in Algorithm~\ref{alg:cluster-adaptation}, the process begins with the initial cluster partition $\mathcal{P}_{0}$ during the prefill phase. For each newly generated KV entry $\mathbf{k}_{\text{new}}^{(t)}$, \DynaKV{} compares it against existing cluster representatives and assigns it to the nearest cluster based on feature similarity. Importantly, \DynaKV{} maintains and updates the variance of each cluster, which can be quickly computed from the centroid, the variance, and the new entry.

These variances serve as effectiveness scores for the clusters, capturing the similarity among KV entries within each cluster. If the variance remains below an acceptable threshold, the cluster is updated to incorporate the new entry. When the variance exceeds the threshold, indicating that the cluster is no longer cohesive, the cluster is flagged for splitting. Since cluster sizes are relatively small (naturally bounded by the variance), the computational cost of splitting is negligible. To account for variations in KVCache distributions across attention heads, the thresholds are defined in a head-specific manner. This scoring mechanism preserves clustering effectiveness (\textit{Objective 1}) while keeping cluster granularity as coarse as possible (\textit{Objective 2}).

\noindent\textbf{Delayed-Split Strategy.} Because flash memory lacks compute capability, cluster splitting must be performed in DRAM, where a local clustering algorithm is executed. In other words, if a cluster resides in flash memory, it must first be transferred to DRAM. To reduce the I/O overhead caused by frequent splits, \DynaKV{} adopts a delayed-split strategy.

Specifically, \DynaKV{} performs actual cluster splitting only when the flagged cluster has been retrieved into DRAM. If the flagged cluster is not yet in memory, the split is deferred. Instead, newly added KV entries belonging to the flagged cluster are stored in a memory buffer until the cluster is eventually loaded. To control memory complexity, a maximum buffer size is pre-defined. Once the buffer is full, \DynaKV{} prioritizes uploading the cluster with the largest number of buffered entries and performs the split operation. Because all deferred entries are guaranteed to be incorporated once the cluster is loaded, this strategy preserves correctness while greatly reducing I/O overhead (\textit{Objective 3}).

\begin{figure}[t]
    \centering
    \includegraphics[width=1.0\linewidth]{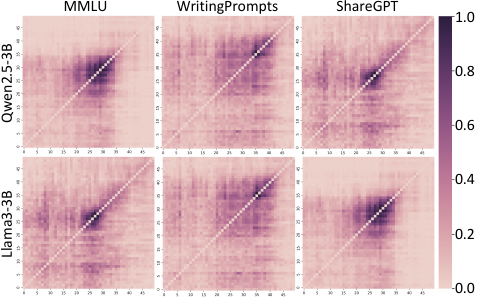}
    \caption{Visualization of inter-cluster correlation across models and datasets. Each matrix element $(i, j)$ denotes the frequency with which cluster $C_{i}$ and $C_{j}$ are co-retrieved.}
    \label{fig:cluster-correlation}
\end{figure}

\section{Continuity-Centric Flash Management}\label{sec:flash}
\subsection{Correlation-Aware Placement}
Given the sparse nature of KVCache retrieval, the KV entries within a cluster are scattered according to strict sequence order. Therefore, transferring a KVCache cluster from flash incurs numerous small-grained read accesses, leaving the I/O operations heavily IOPS-bound on smartphones. To address this issue, \DynaKV{} jointly exploits intra- and inter-cluster correlations to guide data placement on flash, consolidating fragmented accesses and improving I/O efficiency.

\noindent\textbf{Intra-Cluster Correlation.} Since clusters serve as the fundamental retrieval units, all KV entries within a cluster are fetched from flash to DRAM together. To exploit this property, \DynaKV{} adopts a cluster-based placement strategy, arranging KV entries within each cluster continuously on flash rather than preserving their strict sequence order.

\noindent\textbf{Inter-Cluster Correlation.} \DynaKV{} further exploits the frequent co-retrieved patterns observed across clusters. As shown in Figure~\ref{fig:cluster-correlation}, certain clusters are repeatedly accessed together during inference, exhibiting strong inter-cluster correlation. To take advantage of this property, \DynaKV{} constructs an adjacency matrix that records retrieval frequencies between clusters. This step is performed only once before decoding, using the initial input as a reference. By estimating probability using frequency $f$, \DynaKV{} computes the co-retrieval probability of cluster $C_{i}$ and $C_{j}$ as follows:
\begin{equation}
    P(ij)=\frac{f(C_{i},C_{j})}{\sum_{k=1}^{M_{0}}\sum_{l=1}^{M_{0}}f(C_{k},C_{l})}
\end{equation}
where $M_{0}$ denotes the number of initial clusters. Guided by these patterns, \DynaKV{} co-locates highly correlated clusters adjacently in flash. This placement not only preserves intra-cluster locality but also enables multiple clusters to be retrieved through a single sequential access.

\begin{figure}[t]
    \centering
    \includegraphics[width=1.0\linewidth]{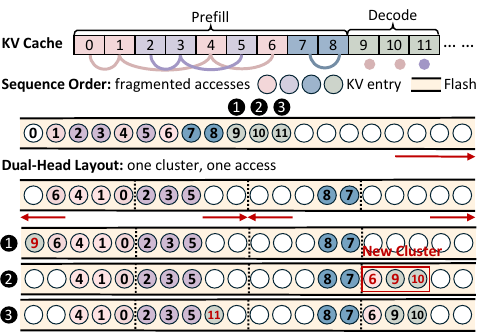}
    \caption{Dual-Head flash layout in \DynaKV{}. Unlike strict sequence order, this layout supports continuous flash access and frequent cluster updates with minimal storage overhead.}
    \label{fig:flash-management}
\end{figure}

\subsection{Dual-Head Cluster Layout}
While correlation-aware placement substantially alleviates the IOPS bottleneck of cluster transfers on smartphones, it introduces a new challenge for efficient cluster updates. Due to the adaptive clustering algorithm in \DynaKV{}, a cluster may either append new entries or be split into two new clusters. Under a strictly contiguous layout, these updates would require costly data movement on flash, significantly degrading inference performance. To overcome this, \DynaKV{} introduces a dual-head cluster layout, placing two correlated clusters in a back-to-back manner. In addition, a page-aligned buffer is employed to eliminate write amplification during updates, further improving the efficiency of \DynaKV{}.

\noindent\textbf{Back-to-Back Manner.} As illustrated in Figure~\ref{fig:flash-management}, \DynaKV{} allocates a dedicated memory pool on flash for each pair of co-retrieved clusters, with pool size set to twice the maximum cluster size. The two clusters are placed back-to-back, growing inward from opposite ends of the pool. This layout enables new entries to be appended seamlessly to either cluster without requiring expensive data permutations. When a cluster is split, one of the resulting sub-clusters continues to grow from the original head, while the other is migrated to a newly allocated pool, paired with another cluster (or left with an empty partner slot). In this way, \DynaKV{} preserves co-retrieval benefits while avoiding costly data relocation.

\noindent\textbf{Page-Aligned Buffer.} Flash writes must align to page boundaries, and small updates often trigger partial-page rewrites, amplifying write traffic. To improve the efficiency of cluster updates, \DynaKV{} employs a page-aligned buffer in memory that aggregates new entries until a full page can be written. Buffers are maintained at the cluster level to isolate updates across different clusters. To conserve memory, buffers are allocated only for hot clusters with frequent updates, while cold clusters issue direct writes when necessary. This design eliminates write amplification during appends and splits while preserving sequential write patterns.

\begin{figure}[t]
    \centering
    \includegraphics[width=1.0\linewidth]{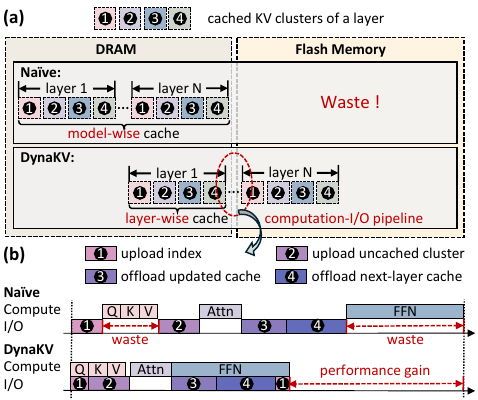}
    \caption{(a) A large virtual cache space is abstracted across DRAM and flash. (b) The additional data transfer overhead is effectively hidden by pipelining with computation.}
    \label{fig:cache-design}
\end{figure}

\section{Memory-Efficient Cache Design}\label{sec:cache}
\subsection{Pipeline-Driven Space Virtualization}
Caching frequently accessed KV entries (including cluster representatives) in memory is critical for reducing I/O overhead, but the limited DRAM capacity on smartphones makes it infeasible to store the full KVCache. To address this limitation, \DynaKV{} introduces a pipeline-driven virtualization mechanism that leverages both DRAM and flash, overlapping cache transfers with computation to hide I/O latency. By effectively virtualizing DRAM capacity, \DynaKV{} provides the abstraction of a large cache space while physically occupying a small memory footprint. As shown in Figure~\ref{fig:cache-design}, this virtualization is realized with two complementary pipelines:

\noindent\textbf{Pipeline 1: Projection ($QKV$).} In KVCache retrieval, cluster representatives are necessary for determining which clusters to access and are typically cached persistently in memory. \DynaKV{} eliminates this static memory cost by storing representatives on flash and pipelining their transfer with query ($Q$) projection. Since representatives are much smaller than full KV entries, most of the transfer time is effectively hidden. Once the representatives are resident in memory, the subsequent KV retrieval can proceed without delay. The transfer of retrieved KV entries is further overlapped with the computation of key-value ($KV$) projection and $Q$-$K$ similarity.

\noindent\textbf{Pipeline 2: Attention and FFN.} Beyond cluster representatives, frequently retrieved KV clusters are also typically cached. Because their size is much larger, overlapping their transfer requires more computation time to fully hide I/O latency. \DynaKV{} addresses this by pipelining cache transfers with computation during the attention and feed-forward network (FFN) stages. While the current layer executes, its KVCache is offloaded to flash, and the KVCache of the next layer is prefetched into DRAM. This layer-wise scheduling ensures that the cache of each layer is available exactly when needed, while unused layers remain virtualized on flash. As a result, memory consumption is bounded at the granularity of a single layer rather than the entire model, greatly improving efficiency under tight DRAM budgets.

\begin{table}[t]
    \small
    \centering
    \caption{Probability of updated clusters being retrieved on MMLU~\cite{mmlu} across LLMs and subsequent decoding steps $t$.}
    \label{tab:kvcache-pattern}
    \begin{adjustbox}{width=1.0\linewidth,center}
    \begin{tabular}{lccccc}
        \toprule
        \textbf{Step (tokens)}      & $t=1$  & $t=2$  & $t=4$  & $t=8$  & $t=16$ \\ \midrule
        Qwen2.5-1.5B~\cite{qwen2.5} & 42.4\% & 47.9\% & 53.7\% & 60.5\% & 67.0\% \\
        Qwen2.5-3B~\cite{qwen2.5}   & 59.4\% & 65.5\% & 70.9\% & 76.3\% & 80.1\% \\
        Llama3.2-1B~\cite{llama3.2} & 36.5\% & 42.3\% & 48.7\% & 54.1\% & 61.4\% \\
        Llama3.2-3B~\cite{llama3.2} & 30.1\% & 36.3\% & 42.8\% & 49.6\% & 54.9\% \\
        \bottomrule
    \end{tabular}
    \end{adjustbox}
\end{table}

\subsection{Cluster-Aligned Replacement Policy}
Beyond expanding cache space, \DynaKV{} further optimizes its cache replacement strategy to maximize utilization under strict memory constraints. Conventional cache replacement policies~\cite{cache-x3,cache-lru,cache-lrfu} (e.g., LRU, LFU) are designed for entry-level access patterns and do not align well with the semantics of cluster-based retrieval. When applied directly to \DynaKV{}, such policies often lead to excessive memory consumption or degraded hit rates. To overcome these limitations, \DynaKV{} introduces a cluster-aligned cache replacement policy that explicitly incorporates the structure and dynamics of cluster-based KV retrieval. The design is guided by two principles:

\noindent\textbf{Principle 1: Prioritize Small Clusters.} Caching large clusters is inefficient for two reasons. First, large clusters occupy disproportionate memory capacity, yet their performance benefit is limited since they already benefit from sequential placement on flash. Second, large clusters are more likely to be split under the adaptive clustering algorithm, which undermines cache stability and can quickly invalidate cached data. In contrast, small clusters consume far less memory and yield higher marginal gains in access efficiency. By prioritizing small clusters, \DynaKV{} reduces memory overhead and maximizes the performance benefit of caching.

\noindent\textbf{Principle 2: Retain Updated Clusters.} \DynaKV{} leverages the unique access patterns of KVCache clusters to further reduce I/O transfers. As presented in Table~\ref{tab:kvcache-pattern}, clusters that have been recently updated—either by appending new entries or by being split into sub-clusters—tend to exhibit stronger short-term relevance. To exploit this property, \DynaKV{} reserves dedicated cache space for updated clusters and retains them in memory regardless of general replacement decisions. This targeted retention capitalizes on the unique temporal locality of KVCache, preventing premature eviction of high-utility clusters and increasing cache hit rates.

\section{Implementation and Discussion}
\noindent\textbf{System Implementation.} We implement \DynaKV{} as an end-to-end system built on \textit{llama.cpp}~\cite{llama-cpp}, the most widely used inference framework for mobile devices. Following prior studies that demonstrate the better performance of CPUs over NPUs for decoding on smartphones~\cite{bitnet,t-mac}, we adopt the CPU as the primary backend. To support the core techniques of \DynaKV{}, we extend the runtime by rewriting the KVCache swapping, flash loading, and cache management modules. For high-performance asynchronous data transfers, we leverage the Linux kernel interface \textit{io\_uring}~\cite{liburing}. To decouple I/O from computation, we create a dedicated data-loading thread using $\textit{ggml\_thread\_create}$, and synchronize it with computation through atomic semaphores. On the computation side, we employ an optimized K-Means algorithm for multi-threaded execution and are robust to empty-cluster cases. We also modify the original \textit{flash\_attention} kernel to handle potential numerical overflow errors introduced by reordered KV computations. Overall, \DynaKV{} introduces 17 new operators and over 5,500 lines of new C++ code.

\noindent\textbf{Discussion 1: Long-Sequence Prefill.} \DynaKV{} is fully compatible with long-sequence prefill workloads. Similar to long-sequence decoding, prefill also involves long-context processing and therefore requires efficient KVCache retrieval on smartphones. However, since the prefill phase dominates the overall sequence length, the impact of KVCache distribution shifts is relatively limited. In such cases, \DynaKV{} behaves comparably to existing cluster-based approaches, with its additional mechanisms introducing no extra overhead. Importantly, although the KVCache generated during prefill is eventually offloaded, the prefill stage itself can incur significant peak memory usage, an issue largely overlooked in prior work. In contrast, the adaptive KVCache management of \DynaKV{} can be applied continuously throughout prefill, reducing memory peak while keeping accuracy.

\noindent\textbf{Discussion 2: CPU-GPU Platform.} Although \DynaKV{} primarily targets CPU-only execution on smartphones, the design naturally extends to heterogeneous CPU–GPU platforms. In this setting, GPU memory acts as the active working space for inference, while CPU memory serves as a staging buffer for KVCache offloading. Similarly, the bandwidth between CPU and GPU memory can become a critical bottleneck, making adaptive KVCache management in \DynaKV{} equally important. In addition, lightweight operations (e.g., clustering updates or variance tracking) can be executed directly on the CPU without I/O communication. This division of workloads makes the integration of \DynaKV{} into heterogeneous CPU–GPU platforms even more straightforward. When CPU storage is also incorporated, \DynaKV{} can finally generalize into a three-tier memory hierarchy, efficiently spanning CPU storage, CPU memory, and GPU memory.

\begin{table}[t]
    \small
    \centering
    \caption{Smartphone hardware configurations.}
    \label{tab:exp-configuration-hardware}
    \begin{adjustbox}{width=1.0\linewidth,center}
    \setlength{\tabcolsep}{1.5pt}
    \begin{tabular}{lllll}
        \toprule
        \textbf{Device}  & \textbf{SoC}        & \textbf{DRAM} & \textbf{Flash} & \textbf{Storage} \\ \midrule
        OnePlus Ace5 Pro & Snapdragon 8 Elite  & 16GB          & 512GB          & UFS4.0           \\
        OnePlus 12       & Snapdragon 8 Gen 3  & 24GB          & 1TB            & UFS4.0           \\
        OnePlus Ace3     & Snapdragon 8 Gen 2  & 16GB          & 512GB          & UFS4.0           \\
        OnePlus Ace2     & Snapdragon 8+ Gen 1 & 16GB          & 512GB          & UFS3.1           \\
        \bottomrule
    \end{tabular}
    \end{adjustbox}
\end{table}

\begin{table}[t]
    \small
    \centering
    \caption{Model configurations.}
    \label{tab:exp-configuration-model}
    \begin{adjustbox}{width=1.0\linewidth,center}
    \setlength{\tabcolsep}{2pt}
    \begin{tabular}{lcccc}
        \toprule
        \textbf{Model}               & \textbf{Size} & \textbf{Attention} & \textbf{Hidden Size} & \textbf{Group Size} \\ \midrule
        Qwen(2.5)-S~\cite{qwen2.5}   & 1.5B          & GQA                & 1,536                & 12/2=6              \\
        Qwen(2.5)-M~\cite{qwen2.5}   & 3.1B          & GQA                & 2,048                & 16/2=8              \\
        Llama(3.2)-S~\cite{llama3.2} & 1.2B          & GQA                & 2,048                & 32/8=4              \\
        Llama(3.2)-M~\cite{llama3.2} & 3.2B          & GQA                & 3,072                & 24/8=3              \\
        Mistral~\cite{mistral}       & 7.8B          & GQA                & 4,096                & 32/8=4              \\
        \bottomrule
    \end{tabular}
    \end{adjustbox}
\end{table}

\begin{figure*}[t]
    \centering
    \includegraphics[width=1.0\linewidth]{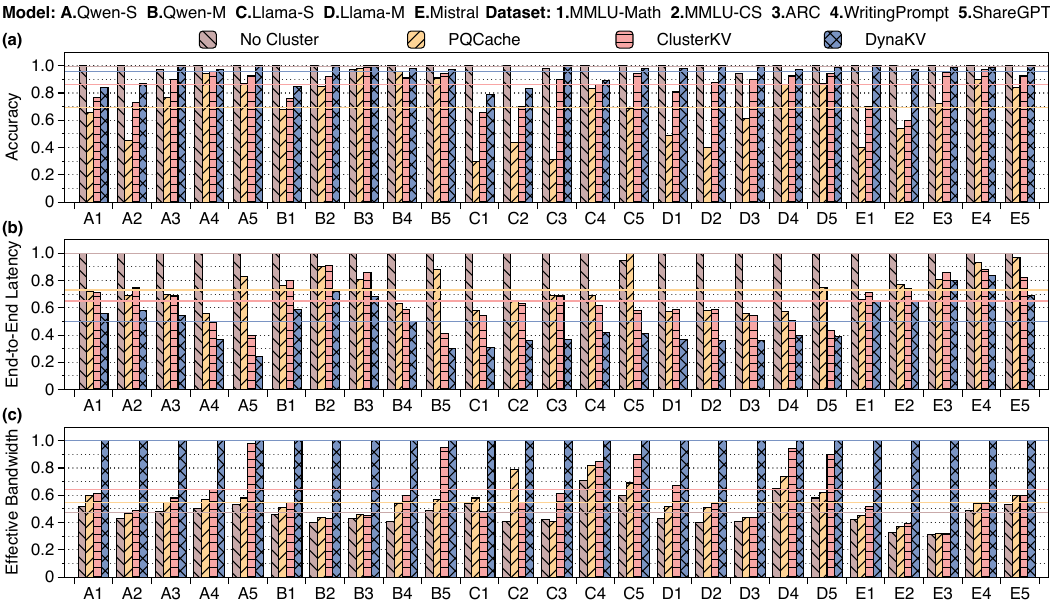}
    \caption{Overall performance (normalized) across various models and datasets on OnePlus 12. Lines indicate average values.}
    \label{fig:exp-overall-performance}
\end{figure*}

\section{Evaluation}
\subsection{Experimental Setup}
\noindent\textbf{Hardware.} We conduct evaluations on four representative smartphones, as detailed in Table~\ref{tab:exp-configuration-hardware}. These devices feature varying SoC configurations, DRAM and flash capacities, and UFS versions, spanning from low-tier to high-tier hardware. All experiments are conducted on Termux~\cite{termux}, an Android shell application, to ensure compatibility across platforms.

\noindent\textbf{Models.} We choose five widely adopted on-device LLMs for evaluation, as listed in Table~\ref{tab:exp-configuration-model}. These models differ in both architecture and parameter scale. Following common practice, all models adopt group-query attention (GQA)~\cite{gqa} with varying group sizes, where larger groups imply greater KVCache reuse. All experiments are conducted in FP8 precision.

\noindent\textbf{Datasets.} We evaluate \DynaKV{} across a diverse set of datasets covering three representative long-sequence decoding scenarios (1) \textit{Reasoning}. We use mathematics and computer science tasks from MMLU\cite{mmlu}, along with the AI2 Reasoning Challenge (ARC)\cite{arc}, a benchmark for model reasoning ability. (2) \textit{Story Generation}. We use Writing Prompts~\cite{writing-prompts}, a large-scale dataset for long-form creative story generation. (3) \textit{Multi-turn Conversion}. We use ShareGPT~\cite{sharegpt}, which contains real multi-turn conversations collected from ChatGPT.

\noindent\textbf{Baselines.} We compare \DynaKV{} against two state-of-the-art KVCache retrieval methods: PQCache~\cite{pqcache} and ClusterKV~\cite{clusterkv}. These represent the two dominant directions for cluster adaptation: \textit{static update} and \textit{local update}, respectively. In addition, we include a naive baseline that performs KVCache retrieval without clustering, to demonstrate the necessity and effectiveness of cluster-based organization.

\noindent\textbf{Metrics.} We evaluate \DynaKV{} using two primary metrics: accuracy and end-to-end latency. For reasoning tasks, accuracy is measured directly on labels. For story generation and multi-turn conversation, we follow common practice~\cite{llm-as-a-judge} and use LLM-based evaluation from multiple perspectives. We normalize the values when large discrepancies occur for clarity. All reported results are averaged over 10 trials.

\subsection{Overall Performance}

\noindent\textbf{Accuracy Evaluation.} Figure~\ref{fig:exp-overall-performance}(a) shows that \DynaKV{} yields the highest accuracy over both cluster-based methods across all models and datasets, approaching the no-cluster baseline (the theoretical upper bound). This demonstrates that \DynaKV{} effectively mitigates the KVCache distribution shift. In contrast, PQCache relies on static updates that fail to adapt to the shift, yielding only 0.69 accuracy on average.

\noindent\textbf{End-to-End Latency.} Figure~\ref{fig:exp-overall-performance}(b) shows that \DynaKV{} yields average speedup improvements of $2.00\times$, $1.47\times$, $1.31\times$ over the no-cluster baseline, PQCache, and ClusterKV, respectively. The no-cluster baseline exhibits the highest latency, underscoring the necessity of clustering. Moreover, the results demonstrate that adaptive KVCache clustering in \DynaKV{} introduces negligible overhead, while system-level I/O optimizations play a dominant role in reducing latency.

\noindent\textbf{Effective Bandwidth.} Figure~\ref{fig:exp-overall-performance}(c) shows that \DynaKV{} delivers average effective bandwidth improvements of $2.10\times$, $1.83\times$, $1.56\times$ over the three baselines. These gains stem from two factors: (1) \DynaKV{} forms more effective clusters that contain fewer redundant KV entries. (2) It alleviates the IOPS bottleneck by reducing fine-grained flash accesses.

\begin{table}[b]
    \small
    \centering
    \caption{Cluster variance across models and datasets (\textit{A-D}: Qwen-S and -M, Llama-S and -M; \textit{1-2}: MMLU-Math and -CS).}
    \label{tab:exp-ablation-cluster-variance}
    \begin{adjustbox}{width=1.0\linewidth,center}
    \setlength{\tabcolsep}{3pt}
    \begin{tabular}{lllllllll}
        \toprule
        \textbf{Cases} & \textbf{A1} & \textbf{A2} & \textbf{B1} & \textbf{B2} & \textbf{C1} & \textbf{C2} & \textbf{D1} & \textbf{D2} \\ \midrule
        PQCache        & 206.7       & 198.2       & 211.3       & 214.2       & 146.1       & 154.4       & 218.1       & 201.6       \\
        ClusterKV      & 122.4       & 125.5       & 124.3       & 127.7       & 95.1        & 94.7        & 129.0       & 132.1       \\
        DynaKV          & 119.6       & 122.5       & 118.3       & 125.2       & 94.6        & 90.8        & 123.3       & 123.8       \\
        \bottomrule
    \end{tabular}
    \end{adjustbox}
\end{table}

\subsection{Ablation Study}
\noindent\textbf{Technique 1: Cluster Adaptation.} Table~\ref{tab:exp-ablation-cluster-variance} shows that clusters in \DynaKV{} exhibit significantly lower variance compared to the baselines. This demonstrates that the variance-based scoring mechanism in \DynaKV{} effectively preserves clustering quality, which is essential for retrieval accuracy.

\begin{figure}
    \centering
    \includegraphics[width=1.0\linewidth]{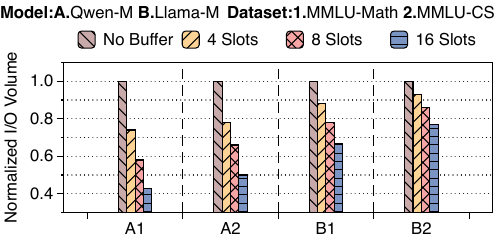}
    \caption{KVCache transfer volume under varying buffer size (for delayed splitting) across models and datasets.}
    \label{fig:exp-ablation-cluster-overhead}
\end{figure}

Figure~\ref{fig:exp-ablation-cluster-overhead} further evaluates the KVCache transfer reduction achieved by the delayed-split strategy. By temporarily retaining a small number of KV entries (up to 16) in memory and deferring splitting until the corresponding cluster is retrieved, \DynaKV{} reduces I/O volume by up to $2.32\times$.

\begin{figure}
    \centering
    \includegraphics[width=1.0\linewidth]{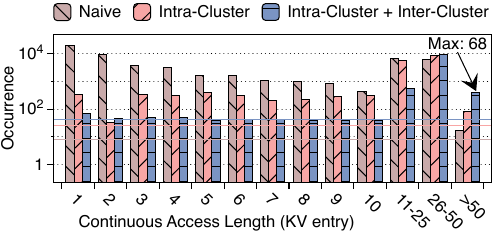}
    \caption{Statistical information on flash access lengths of Qwen-M on MMLU. Lines indicate average values.}
    \label{fig:exp-ablation-flash-continuity}
\end{figure}

\noindent\textbf{Technique 2: Flash Management.} Figure~\ref{fig:exp-ablation-flash-continuity} demonstrates the effectiveness of both intra-cluster and inter-cluster continuity designs in \DynaKV{}. Under strict sequence order, the average flash access length is limited to below 7.9 entries. By exploiting cluster correlations, \DynaKV{} increases continuous access lengths to 25.3 and 40.8, respectively. Notably, the maximum continuous access length reaches 68 entries.

\begin{figure}
    \centering
    \includegraphics[width=1.0\linewidth]{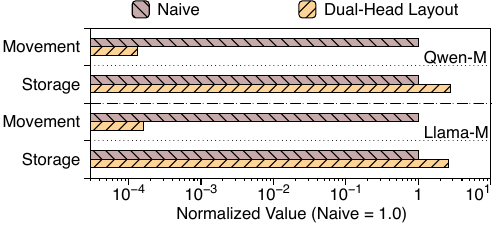}
    \caption{Data movement and storage usage with and without the dual-head layout across different models.}
    \label{fig:exp-ablation-flash-movement}
\end{figure}

Figure~\ref{fig:exp-ablation-flash-movement} shows the effectiveness of our dual-head flash layout. With only minor overhead in abundant storage, this design drastically reduces costly data permutation. Compared to the naive cluster-by-cluster layout, \DynaKV{} achieves over $10^4$ reduction in data movement across both models.

\begin{figure}
    \centering
    \includegraphics[width=1.0\linewidth]{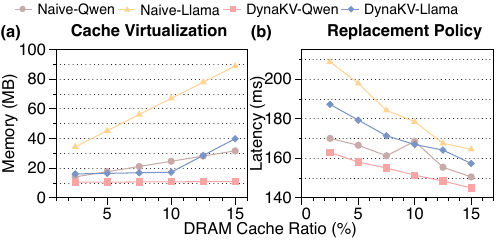}
    \caption{(a) Memory usage and (b) End-to-end latency under varying cache ratios with and without specific design.}
    \label{fig:exp-ablation-cache}
\end{figure}

\noindent\textbf{Technique 3: Cache Design.} Figure~\ref{fig:exp-ablation-cache}(a) demonstrates that \DynaKV{} reduces cache memory usage across different cache ratios. By pipelining data transfer for the next layer with ongoing computation, cache virtualization shifts memory demand from the model level to the layer level, yielding increasing savings as the cache ratio grows (up to $2.22\times$).

Figure~\ref{fig:exp-ablation-cache}(b) shows that \DynaKV{} also outperforms the baselines in latency across different cache ratios. Compared with standard LRU, its cluster-aligned replacement policy achieves a higher cache hit rate under the same cache size, reducing I/O transfers and delivering up to $1.12\times$ speedups.

\begin{table}[b]
    \small
    \centering
    \caption{End-to-end latency under varying decoding lengths on MMLU across different models using OnePlus 12.}
    \label{tab:exp-sensitivity-length}
    \begin{adjustbox}{width=1.0\linewidth,center}
    \setlength{\tabcolsep}{3pt}
    \begin{tabular}{lccccc}
        \toprule
        \textbf{Length} & $1K$     & $2K$     & $4K$     & $8K$      & $16K$     \\ \midrule
        Qwen-S          & 84.6 ms  & 96.5 ms  & 123.5 ms & 209.2 ms  & 376.3 ms  \\
        Qwen-M          & 152.5 ms & 167.5 ms & 215.1 ms & 326.3 ms  & 636.8 ms  \\
        Llama-S         & 76.7 ms  & 97.6 ms  & 189.4 ms & 346.7 ms  & 677.3 ms  \\
        Llama-M         & 189.3 ms & 289.6 ms & 499.6 ms & 1044.0 ms & 1802.3 ms \\
        \bottomrule
    \end{tabular}
    \end{adjustbox}
\end{table}

\subsection{Sensitivity Analysis}
\noindent\textbf{Impact of Decoding Lengths.} We first evaluate the performance of \DynaKV{} across decoding lengths in long-sequence decoding, as listed in Table~\ref{tab:exp-sensitivity-length}. The results show that performance scales nearly linearly with sequence length up to $16K$, indicating the strong scalability of \DynaKV{} to long contexts without incurring notable overhead from cluster adaptation.

\begin{figure}
    \centering
    \includegraphics[width=1.0\linewidth]{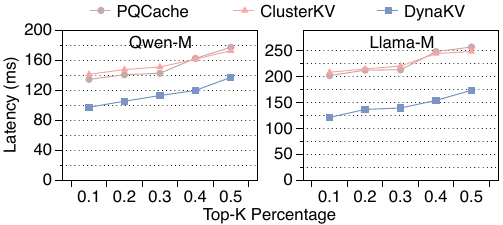}
    \caption{End-to-end latency under varying top-$k$ percentages of KVCache retrieval on MMLU using OnePlus 12.}
    \label{fig:exp-sensitivity-sparsity}
\end{figure}

\noindent\textbf{Impact of Sparsity Levels.} We next evaluate how variations in KVCache sparsity levels affect the performance of \DynaKV{}, as shown in Figure~\ref{fig:exp-sensitivity-sparsity}. The results show that \DynaKV{} consistently outperforms the baselines across different top-$k$ retrieval ratios. The performance gains are more pronounced at higher sparsity levels, where accurate KVCache clustering and IOPS optimizations become more critical.

\begin{figure}
    \centering
    \includegraphics[width=1.0\linewidth]{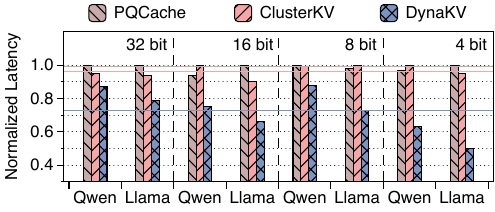}
    \caption{End-to-end latency of Qwen-M and Llama-M on MMLU across parameter precisions using OnePlus 12.}
    \label{fig:exp-sensitivity-precision}
\end{figure}

\noindent\textbf{Impact of Precision.} Figure~\ref{fig:exp-sensitivity-precision} shows that \DynaKV{} scales efficiently across different precisions, sustaining an average speedup of $1.36\times$ over two models. Notably, the benefits of \DynaKV{} are even more pronounced when LLMs are quantized from 8-bit to 4-bit precision, since smaller neuron sizes exacerbate the negative impact of scattered flash access.

\begin{figure}
    \centering
    \includegraphics[width=1.0\linewidth]{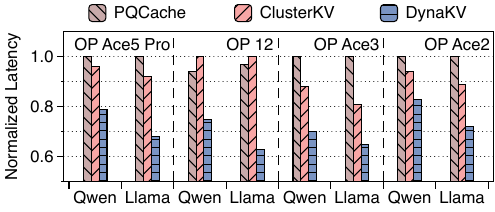}
    \caption{End-to-end latency of Qwen-M and Llama-M on MMLU across diverse hardware configurations.}
    \label{fig:exp-sensitivity-hardware}
\end{figure}

\noindent\textbf{Impact of Hardware.} Figure~\ref{fig:exp-sensitivity-hardware} shows that \DynaKV{} consistently outperforms both baselines across diverse hardware configurations, achieving average $1.28\times$ and $1.48\times$ speedups across the two models. Among the tested devices, the OnePlus Ace2 exhibits the lowest performance due to its weaker UFS, making it easier to reach bandwidth limitations.

\begin{figure}
    \centering
    \includegraphics[width=1.0\linewidth]{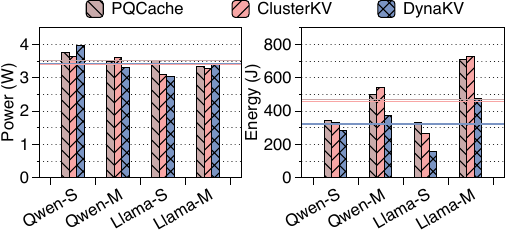}
    \caption{Power and energy consumption across models on MMLU using OnePlus 12. Lines indicate average values.}
    \label{fig:exp-sensitivity-power}
\end{figure}

\noindent\textbf{Impact of Energy Consumption.} Figure~\ref{fig:exp-sensitivity-power} shows that \DynaKV{} draws nearly the same power as the two baselines, since cluster splitting incurs minimal computation overhead. However, thanks to its latency improvements, \DynaKV{} lowers the overall energy consumption by $1.57\times$ on average.

\section{Related Works}
\noindent\textbf{On-Device LLM Inference.} With growing concerns over privacy regulations and the need for real-time response, on-device LLM inference has become increasingly important. Prior efforts have largely focused on two directions: improving computational efficiency~\cite{t-mac,heterollm,mllm} and alleviating memory bottlenecks by offloading model weights to flash storage~\cite{neuralink,activeflow,powerinfer2,llm-flash}. \DynaKV{} extends this second line of work to long-sequence decoding, where the KVCache, rather than model weights, emerges as the dominant memory bottleneck.

\noindent\textbf{KVCache Compression.} A promising direction for alleviating the KVCache memory bottleneck is compression. Some methods~\cite{streamingllm,h2o,scissorhands,fastgen} selectively retain only the most important tokens while evicting less important ones, thereby reducing memory usage. Others~\cite{kvquant,flexgen,atom,llm-qat,kivi} focus on quantizing KVCache activations to lower precision without sacrificing accuracy. \DynaKV{} is orthogonal to these methods and can be combined with them for further gains.

\noindent\textbf{KVCache Retrieval.} Recent works~\cite{quest,snapkv,sparq-attention,infllm,retrieval-attention} exploit the inherent sparsity of the attention mechanism to reduce KVCache memory by retrieving only query-relevant tokens. To further improve retrieval efficiency, other approaches~\cite{pqcache,clusterkv,chelsea,squeezed-attention,shadowkv} organize the KVCache into clusters for indexing. However, all these methods overlook the distribution shift of the KVCache during decoding, leading to potential accuracy degradation. Moreover, they lack optimizations tailored for mobile devices, where stringent memory hierarchy constraints make sustaining efficiency especially challenging. \DynaKV{} explicitly tackles both issues.

\section{Conclusion}
This paper introduces \DynaKV{}, the first adaptive KVCache management approach designed to jointly address accuracy and efficiency for long-sequence LLM decoding on smartphones. Building on algorithmic innovations in adaptive KVCache clustering, \DynaKV{} incorporates system-level optimizations that span the entire memory hierarchy. We envision \DynaKV{} bringing us closer to realizing day-long intelligent agents on mobile devices, capable of continuously assisting people in their daily lives.

\bibliographystyle{plain}
\bibliography{references}

\begin{thebibliography}{10}

\bibitem{gqa}
Joshua Ainslie, James Lee-Thorp, Michiel De~Jong, Yury Zemlyanskiy, Federico Lebr{\'o}n, and Sumit Sanghai.
\newblock Gqa: Training generalized multi-query transformer models from multi-head checkpoints.
\newblock {\em arXiv preprint arXiv:2305.13245}, 2023.

\bibitem{llm-flash}
Keivan Alizadeh, Seyed~Iman Mirzadeh, Dmitry Belenko, S~Khatamifard, Minsik Cho, Carlo~C Del~Mundo, Mohammad Rastegari, and Mehrdad Farajtabar.
\newblock Llm in a flash: Efficient large language model inference with limited memory.
\newblock In {\em Proceedings of the 62nd Annual Meeting of the Association for Computational Linguistics (Volume 1: Long Papers)}, pages 12562--12584, 2024.

\bibitem{anns-1}
Sunil Arya, David~M Mount, Nathan~S Netanyahu, Ruth Silverman, and Angela~Y Wu.
\newblock An optimal algorithm for approximate nearest neighbor searching fixed dimensions.
\newblock {\em Journal of the ACM (JACM)}, 45(6):891--923, 1998.

\bibitem{liburing}
Jens Axboe.
\newblock liburing.
\newblock \url{https://github.com/axboe/liburing}, 2025.
\newblock Accessed: 2025-02-25.

\bibitem{mt-bench-101}
Ge~Bai, Jie Liu, Xingyuan Bu, Yancheng He, Jiaheng Liu, Zhanhui Zhou, Zhuoran Lin, Wenbo Su, Tiezheng Ge, Bo~Zheng, et~al.
\newblock Mt-bench-101: A fine-grained benchmark for evaluating large language models in multi-turn dialogues.
\newblock {\em arXiv preprint arXiv:2402.14762}, 2024.

\bibitem{heterollm}
Le~Chen, Dahu Feng, Erhu Feng, Rong Zhao, Yingrui Wang, Yubin Xia, Haibo Chen, and Pinjie Xu.
\newblock Heterollm: Accelerating large language model inference on mobile socs platform with heterogeneous ai accelerators.
\newblock {\em arXiv preprint arXiv:2501.14794}, 2025.

\bibitem{document-1}
Robert Chew, John Bollenbacher, Michael Wenger, Jessica Speer, and Annice Kim.
\newblock Llm-assisted content analysis: Using large language models to support deductive coding.
\newblock {\em arXiv preprint arXiv:2306.14924}, 2023.

\bibitem{arc}
Peter Clark, Isaac Cowhey, Oren Etzioni, Tushar Khot, Ashish Sabharwal, Carissa Schoenick, and Oyvind Tafjord.
\newblock Think you have solved question answering? try arc, the ai2 reasoning challenge.
\newblock {\em arXiv preprint arXiv:1803.05457}, 2018.

\bibitem{gemini2.5}
Gheorghe Comanici, Eric Bieber, Mike Schaekermann, Ice Pasupat, Noveen Sachdeva, Inderjit Dhillon, Marcel Blistein, Ori Ram, Dan Zhang, Evan Rosen, et~al.
\newblock Gemini 2.5: Pushing the frontier with advanced reasoning, multimodality, long context, and next generation agentic capabilities.
\newblock {\em arXiv preprint arXiv:2507.06261}, 2025.

\bibitem{attention-sparsity}
Yichuan Deng, Zhao Song, and Chiwun Yang.
\newblock Attention is naturally sparse with gaussian distributed input.
\newblock {\em CoRR}, 2024.

\bibitem{document-2}
Darren Edge, Ha~Trinh, Newman Cheng, Joshua Bradley, Alex Chao, Apurva Mody, Steven Truitt, Dasha Metropolitansky, Robert~Osazuwa Ness, and Jonathan Larson.
\newblock From local to global: A graph rag approach to query-focused summarization.
\newblock {\em arXiv preprint arXiv:2404.16130}, 2024.

\bibitem{writing-prompts}
Angela Fan, Mike Lewis, and Yann Dauphin.
\newblock Hierarchical neural story generation.
\newblock {\em arXiv preprint arXiv:1805.04833}, 2018.

\bibitem{fastgen}
Suyu Ge, Yunan Zhang, Liyuan Liu, Minjia Zhang, Jiawei Han, and Jianfeng Gao.
\newblock Model tells you what to discard: Adaptive kv cache compression for llms.
\newblock {\em arXiv preprint arXiv:2310.01801}, 2023.

\bibitem{llama-cpp}
Georgi Gerganov.
\newblock ggerganov/llama.cpp: Port of facebook’s llama model in c/c++.
\newblock \url{https://github.com/ggerganov/llama.cpp}, 2024.

\bibitem{deepseek-r1}
Daya Guo, Dejian Yang, Haowei Zhang, Junxiao Song, Ruoyu Zhang, Runxin Xu, Qihao Zhu, Shirong Ma, Peiyi Wang, Xiao Bi, et~al.
\newblock Deepseek-r1: Incentivizing reasoning capability in llms via reinforcement learning.
\newblock {\em arXiv preprint arXiv:2501.12948}, 2025.

\bibitem{codebase-1}
Daya Guo, Qihao Zhu, Dejian Yang, Zhenda Xie, Kai Dong, Wentao Zhang, Guanting Chen, Xiao Bi, Yu~Wu, YK~Li, et~al.
\newblock Deepseek-coder: When the large language model meets programming--the rise of code intelligence.
\newblock {\em arXiv preprint arXiv:2401.14196}, 2024.

\bibitem{mmlu}
Dan Hendrycks, Collin Burns, Steven Basart, Andy Zou, Mantas Mazeika, Dawn Song, and Jacob Steinhardt.
\newblock Measuring massive multitask language understanding.
\newblock {\em arXiv preprint arXiv:2009.03300}, 2020.

\bibitem{squeezed-attention}
Coleman Hooper, Sehoon Kim, Hiva Mohammadzadeh, Monishwaran Maheswaran, Sebastian Zhao, June Paik, Michael~W Mahoney, Kurt Keutzer, and Amir Gholami.
\newblock Squeezed attention: Accelerating long context length llm inference.
\newblock {\em arXiv preprint arXiv:2411.09688}, 2024.

\bibitem{kvquant}
Coleman Hooper, Sehoon Kim, Hiva Mohammadzadeh, Michael~W Mahoney, Yakun~S Shao, Kurt Keutzer, and Amir Gholami.
\newblock Kvquant: Towards 10 million context length llm inference with kv cache quantization.
\newblock {\em Advances in Neural Information Processing Systems}, 37:1270--1303, 2024.

\bibitem{chelsea}
Jie Hu, Shengnan Wang, Yutong He, Ping Gong, Jiawei Yi, Juncheng Zhang, Youhui Bai, Renhai Chen, Gong Zhang, Cheng Li, et~al.
\newblock Efficient long-context llm inference via kv cache clustering.
\newblock {\em arXiv preprint arXiv:2506.11418}, 2025.

\bibitem{openai-o1}
Aaron Jaech, Adam Kalai, Adam Lerer, Adam Richardson, Ahmed El-Kishky, Aiden Low, Alec Helyar, Aleksander Madry, Alex Beutel, Alex Carney, et~al.
\newblock Openai o1 system card.
\newblock {\em arXiv preprint arXiv:2412.16720}, 2024.

\bibitem{ufs}
{JEDEC}.
\newblock Jedec announces publication of universal flash storage (ufs) standard.
\newblock \url{https://www.jedec.org}, February 2021.
\newblock Accessed: 2024-10-02.

\bibitem{anns-2}
Herve Jegou, Matthijs Douze, and Cordelia Schmid.
\newblock Product quantization for nearest neighbor search.
\newblock {\em IEEE transactions on pattern analysis and machine intelligence}, 33(1):117--128, 2010.

\bibitem{activeflow}
Fucheng Jia, Zewen Wu, Shiqi Jiang, Huiqiang Jiang, Qianxi Zhang, Yuqing Yang, Yunxin Liu, Ju~Ren, Deyu Zhang, and Ting Cao.
\newblock Scaling up on-device llms via active-weight swapping between dram and flash.
\newblock {\em arXiv preprint arXiv:2504.08378}, 2025.

\bibitem{mistral}
Albert~Q. Jiang, Alexandre Sablayrolles, Arthur Mensch, Chris Bamford, Devendra~Singh Chaplot, Diego de~las Casas, Florian Bressand, Gianna Lengyel, Guillaume Lample, Lucile Saulnier, Lélio~Renard Lavaud, Marie-Anne Lachaux, Pierre Stock, Teven~Le Scao, Thibaut Lavril, Thomas Wang, Timothée Lacroix, and William~El Sayed.
\newblock Mistral 7b, 2023.

\bibitem{cache-x3}
Theodore Johnson and Dennis Shasha.
\newblock X3: A low overhead high performance buffer management replacement algorithm.
\newblock In {\em Proceedings of the 20th VLDB Conference}, pages 439--450, 1994.

\bibitem{longlamp}
Ishita Kumar, Snigdha Viswanathan, Sushrita Yerra, Alireza Salemi, Ryan~A Rossi, Franck Dernoncourt, Hanieh Deilamsalehy, Xiang Chen, Ruiyi Zhang, Shubham Agarwal, et~al.
\newblock Longlamp: A benchmark for personalized long-form text generation.
\newblock {\em arXiv preprint arXiv:2407.11016}, 2024.

\bibitem{cache-lrfu}
Donghee Lee, Jongmoo Choi, Jong-Hun Kim, Sam~H Noh, Sang~Lyul Min, Yookun Cho, and Chong~Sang Kim.
\newblock Lrfu: A spectrum of policies that subsumes the least recently used and least frequently used policies.
\newblock {\em IEEE transactions on Computers}, 50(12):1352--1361, 2001.

\bibitem{snapkv}
Yuhong Li, Yingbing Huang, Bowen Yang, Bharat Venkitesh, Acyr Locatelli, Hanchen Ye, Tianle Cai, Patrick Lewis, and Deming Chen.
\newblock Snapkv: Llm knows what you are looking for before generation.
\newblock {\em Advances in Neural Information Processing Systems}, 37:22947--22970, 2024.

\bibitem{retrieval-attention}
Di~Liu, Meng Chen, Baotong Lu, Huiqiang Jiang, Zhenhua Han, Qianxi Zhang, Qi~Chen, Chengruidong Zhang, Bailu Ding, Kai Zhang, et~al.
\newblock Retrievalattention: Accelerating long-context llm inference via vector retrieval.
\newblock {\em arXiv preprint arXiv:2409.10516}, 2024.

\bibitem{clusterkv}
Guangda Liu, Chengwei Li, Jieru Zhao, Chenqi Zhang, and Minyi Guo.
\newblock Clusterkv: Manipulating llm kv cache in semantic space for recallable compression.
\newblock {\em arXiv preprint arXiv:2412.03213}, 2024.

\bibitem{llm-qat}
Zechun Liu, Barlas Oguz, Changsheng Zhao, Ernie Chang, Pierre Stock, Yashar Mehdad, Yangyang Shi, Raghuraman Krishnamoorthi, and Vikas Chandra.
\newblock Llm-qat: Data-free quantization aware training for large language models.
\newblock {\em arXiv preprint arXiv:2305.17888}, 2023.

\bibitem{scissorhands}
Zichang Liu, Aditya Desai, Fangshuo Liao, Weitao Wang, Victor Xie, Zhaozhuo Xu, Anastasios Kyrillidis, and Anshumali Shrivastava.
\newblock Scissorhands: Exploiting the persistence of importance hypothesis for llm kv cache compression at test time.
\newblock {\em Advances in Neural Information Processing Systems}, 36:52342--52364, 2023.

\bibitem{deja-vu}
Zichang Liu, Jue Wang, Tri Dao, Tianyi Zhou, Binhang Yuan, Zhao Song, Anshumali Shrivastava, Ce~Zhang, Yuandong Tian, Christopher Re, et~al.
\newblock Deja vu: Contextual sparsity for efficient llms at inference time.
\newblock In {\em International Conference on Machine Learning}, pages 22137--22176. PMLR, 2023.

\bibitem{kivi}
Zirui Liu, Jiayi Yuan, Hongye Jin, Shaochen Zhong, Zhaozhuo Xu, Vladimir Braverman, Beidi Chen, and Xia Hu.
\newblock Kivi: A tuning-free asymmetric 2bit quantization for kv cache.
\newblock {\em arXiv preprint arXiv:2402.02750}, 2024.

\bibitem{anns-3}
Yu~A Malkov and Dmitry~A Yashunin.
\newblock Efficient and robust approximate nearest neighbor search using hierarchical navigable small world graphs.
\newblock {\em IEEE transactions on pattern analysis and machine intelligence}, 42(4):824--836, 2018.

\bibitem{llama3.2}
{Meta AI}.
\newblock Llama 3.2: Revolutionizing edge ai and vision with open, customizable models for edge \& mobile devices.
\newblock Blog post, Meta AI, September 2024.

\bibitem{codebase-2}
Daye Nam, Andrew Macvean, Vincent Hellendoorn, Bogdan Vasilescu, and Brad Myers.
\newblock Using an llm to help with code understanding.
\newblock In {\em Proceedings of the IEEE/ACM 46th International Conference on Software Engineering}, pages 1--13, 2024.

\bibitem{cache-lru}
Elizabeth~J O'neil, Patrick~E O'neil, and Gerhard Weikum.
\newblock The lru-k page replacement algorithm for database disk buffering.
\newblock {\em Acm Sigmod Record}, 22(2):297--306, 1993.

\bibitem{hellobench}
Haoran Que, Feiyu Duan, Liqun He, Yutao Mou, Wangchunshu Zhou, Jiaheng Liu, Wenge Rong, Zekun~Moore Wang, Jian Yang, Ge~Zhang, et~al.
\newblock Hellobench: Evaluating long text generation capabilities of large language models.
\newblock {\em arXiv preprint arXiv:2409.16191}, 2024.

\bibitem{qwen2.5}
Qwen, :, An~Yang, Baosong Yang, Beichen Zhang, Binyuan Hui, Bo~Zheng, Bowen Yu, Chengyuan Li, Dayiheng Liu, Fei Huang, Haoran Wei, Huan Lin, Jian Yang, Jianhong Tu, Jianwei Zhang, Jianxin Yang, Jiaxi Yang, Jingren Zhou, Junyang Lin, Kai Dang, Keming Lu, Keqin Bao, Kexin Yang, Le~Yu, Mei Li, Mingfeng Xue, Pei Zhang, Qin Zhu, Rui Men, Runji Lin, Tianhao Li, Tianyi Tang, Tingyu Xia, Xingzhang Ren, Xuancheng Ren, Yang Fan, Yang Su, Yichang Zhang, Yu~Wan, Yuqiong Liu, Zeyu Cui, Zhenru Zhang, and Zihan Qiu.
\newblock Qwen2.5 technical report, 2025.

\bibitem{sparq-attention}
Luka Ribar, Ivan Chelombiev, Luke Hudlass-Galley, Charlie Blake, Carlo Luschi, and Douglas Orr.
\newblock Sparq attention: Bandwidth-efficient llm inference.
\newblock {\em arXiv preprint arXiv:2312.04985}, 2023.

\bibitem{sharegpt}
{RyokoAI}.
\newblock {RyokoAI/ShareGPT52K}: Sharegpt human–ai conversations dataset.
\newblock \url{https://huggingface.co/datasets/RyokoAI/ShareGPT52K}, 2023.
\newblock CC0 1.0 license. Approx.\ 52\,000 (legacy) to 90\,000 conversations from ShareGPT API.

\bibitem{flexgen}
Ying Sheng, Lianmin Zheng, Binhang Yuan, Zhuohan Li, Max Ryabinin, Beidi Chen, Percy Liang, Christopher R{\'e}, Ion Stoica, and Ce~Zhang.
\newblock Flexgen: High-throughput generative inference of large language models with a single gpu.
\newblock In {\em International Conference on Machine Learning}, pages 31094--31116. PMLR, 2023.

\bibitem{shadowkv}
Hanshi Sun, Li-Wen Chang, Wenlei Bao, Size Zheng, Ningxin Zheng, Xin Liu, Harry Dong, Yuejie Chi, and Beidi Chen.
\newblock Shadowkv: Kv cache in shadows for high-throughput long-context llm inference.
\newblock {\em arXiv preprint arXiv:2410.21465}, 2024.

\bibitem{mt-bench++}
Yuchong Sun, Che Liu, Kun Zhou, Jinwen Huang, Ruihua Song, Wayne~Xin Zhao, Fuzheng Zhang, Di~Zhang, and Kun Gai.
\newblock Parrot: Enhancing multi-turn instruction following for large language models.
\newblock {\em arXiv preprint arXiv:2310.07301}, 2023.

\bibitem{quest}
Jiaming Tang, Yilong Zhao, Kan Zhu, Guangxuan Xiao, Baris Kasikci, and Song Han.
\newblock Quest: Query-aware sparsity for efficient long-context llm inference.
\newblock {\em arXiv preprint arXiv:2406.10774}, 2024.

\bibitem{termux}
Termux.
\newblock Termux app.
\newblock \url{https://github.com/termux/termux-app}, 2025.
\newblock Accessed: 2025-02-25.

\bibitem{bitnet}
Jinheng Wang, Hansong Zhou, Ting Song, Shijie Cao, Yan Xia, Ting Cao, Jianyu Wei, Shuming Ma, Hongyu Wang, and Furu Wei.
\newblock Bitnet. cpp: Efficient edge inference for ternary llms.
\newblock {\em arXiv preprint arXiv:2502.11880}, 2025.

\bibitem{jenga}
Tuowei Wang, Xingyu Chen, Kun Li, Ting Cao, Ju~Ren, and Yaoxue Zhang.
\newblock $\{$JENGA$\}$: Enhancing $\{$LLM$\}$$\{$Long-Context$\}$ fine-tuning with contextual token sparsity.
\newblock In {\em 2025 USENIX Annual Technical Conference (USENIX ATC 25)}, pages 123--141, 2025.

\bibitem{neuralink}
Tuowei Wang, Ruwen Fan, Minxing Huang, Zixu Hao, Kun Li, Ting Cao, Youyou Lu, Yaoxue Zhang, and Ju~Ren.
\newblock Neuralink: Fast on-device llm inference with neuron co-activation linking.
\newblock In {\em Proceedings of the 30th ACM International Conference on Architectural Support for Programming Languages and Operating Systems, Volume 3}, pages 147--162, 2025.

\bibitem{long-exposure}
Tuowei Wang, Kun Li, Zixu Hao, Donglin Bai, Ju~Ren, Yaoxue Zhang, Ting Cao, and Mao Yang.
\newblock Long exposure: Accelerating parameter-efficient fine-tuning for llms under shadowy sparsity.
\newblock In {\em SC24: International Conference for High Performance Computing, Networking, Storage and Analysis}, pages 1--18. IEEE, 2024.

\bibitem{t-mac}
Jianyu Wei, Shijie Cao, Ting Cao, Lingxiao Ma, Lei Wang, Yanyong Zhang, and Mao Yang.
\newblock T-mac: Cpu renaissance via table lookup for low-bit llm deployment on edge.
\newblock In {\em Proceedings of the Twentieth European Conference on Computer Systems}, pages 278--292, 2025.

\bibitem{pca}
Svante Wold, Kim Esbensen, and Paul Geladi.
\newblock Principal component analysis.
\newblock {\em Chemometrics and intelligent laboratory systems}, 2(1-3):37--52, 1987.

\bibitem{infllm}
Chaojun Xiao, Pengle Zhang, Xu~Han, Guangxuan Xiao, Yankai Lin, Zhengyan Zhang, Zhiyuan Liu, and Maosong Sun.
\newblock Infllm: Training-free long-context extrapolation for llms with an efficient context memory.
\newblock {\em Advances in Neural Information Processing Systems}, 37:119638--119661, 2024.

\bibitem{streamingllm}
Guangxuan Xiao, Yuandong Tian, Beidi Chen, Song Han, and Mike Lewis.
\newblock Efficient streaming language models with attention sinks.
\newblock {\em arXiv preprint arXiv:2309.17453}, 2023.

\bibitem{mllm}
Daliang Xu, Hao Zhang, Liming Yang, Ruiqi Liu, Gang Huang, Mengwei Xu, and Xuanzhe Liu.
\newblock Fast on-device llm inference with npus.
\newblock 2025.

\bibitem{powerinfer2}
Zhenliang Xue, Yixin Song, Zeyu Mi, Xinrui Zheng, Yubin Xia, and Haibo Chen.
\newblock Powerinfer-2: Fast large language model inference on a smartphone.
\newblock {\em arXiv preprint arXiv:2406.06282}, 2024.

\bibitem{seed-story}
Shuai Yang, Yuying Ge, Yang Li, Yukang Chen, Yixiao Ge, Ying Shan, and Yingcong Chen.
\newblock Seed-story: Multimodal long story generation with large language model.
\newblock {\em arXiv preprint arXiv:2407.08683}, 2024.

\bibitem{pqcache}
Hailin Zhang, Xiaodong Ji, Yilin Chen, Fangcheng Fu, Xupeng Miao, Xiaonan Nie, Weipeng Chen, and Bin Cui.
\newblock Pqcache: Product quantization-based kvcache for long context llm inference.
\newblock {\em Proceedings of the ACM on Management of Data}, 3(3):1--30, 2025.

\bibitem{h2o}
Zhenyu Zhang, Ying Sheng, Tianyi Zhou, Tianlong Chen, Lianmin Zheng, Ruisi Cai, Zhao Song, Yuandong Tian, Christopher R{\'e}, Clark Barrett, et~al.
\newblock H2o: Heavy-hitter oracle for efficient generative inference of large language models.
\newblock {\em Advances in Neural Information Processing Systems}, 36:34661--34710, 2023.

\bibitem{atom}
Yilong Zhao, Chien-Yu Lin, Kan Zhu, Zihao Ye, Lequn Chen, Size Zheng, Luis Ceze, Arvind Krishnamurthy, Tianqi Chen, and Baris Kasikci.
\newblock Atom: Low-bit quantization for efficient and accurate llm serving.
\newblock {\em Proceedings of Machine Learning and Systems}, 6:196--209, 2024.

\bibitem{mt-bench}
Lianmin Zheng, Wei-Lin Chiang, Ying Sheng, Siyuan Zhuang, Zhanghao Wu, Yonghao Zhuang, Zi~Lin, Zhuohan Li, Dacheng Li, Eric Xing, et~al.
\newblock Judging llm-as-a-judge with mt-bench and chatbot arena.
\newblock {\em Advances in neural information processing systems}, 36:46595--46623, 2023.

\bibitem{llm-as-a-judge}
Lianmin Zheng, Wei-Lin Chiang, Ying Sheng, Siyuan Zhuang, Zhanghao Wu, Yonghao Zhuang, Zi~Lin, Zhuohan Li, Dacheng Li, Eric Xing, et~al.
\newblock Judging llm-as-a-judge with mt-bench and chatbot arena.
\newblock {\em Advances in neural information processing systems}, 36:46595--46623, 2023.

\end{thebibliography}

\end{document}